# Automated Extraction and Creation of FBS Design Reasoning Knowledge Graphs from Structured Data in Product Catalogues Lacking Contextual Information

[1]Vijayalaxmi Sahadevan*, [1]Sushil Mario, [1]Yash Jaiswal, [1]Divyanshu Bajpai, [1]Vishal Singh, [2]Hiralal Agarwal, [2]Suhas Suresh, and [2]Manjunath Maigur

[1]Indian Institute of Science Bengaluru, India

[2]Harting India Private Limited, India

*Corresponding Author

## Abstract

Ontology-based knowledge graphs (KG) are desirable for effective knowledge management and reuse in various decision-making scenarios, including design. Creating and populating extensive KG based on specific ontological models can be highly labour and time-intensive unless automated processes are developed for knowledge extraction and graph creation. Most research and development on automated extraction and creation of KG is based on extensive unstructured data sets that provide contextual information. However, some of the most useful information about the products and services of a company has traditionally been recorded as structured data. Such structured data sets rarely follow a standard ontology, do not capture explicit mapping of relationships between the entities, and provide no contextual information. Therefore, this research reports a method and digital workflow developed to address this gap. The developed method and workflow employ rule-based techniques to extract and create a Function-Behaviour-Structure (FBS) ontology-based KG from legacy structured data, especially specification sheets and product catalogues. The solution approach consists of two main components: a process for deriving context and context-based classification rules for FBS ontology concepts and a workflow for populating and retrieving the FBS ontology-based KG. KG and Natural Language Processing (NLP) are used to automate knowledge extraction, representation, and retrieval. The workflow's effectiveness is demonstrated via pilot implementation in an industrial context. Insights gained from the pilot study are reported regarding the challenges and opportunities, including discussing the FBS ontology and concepts.

# Introduction

There is growing interest among enterprises and their knowledge engineers in updating and migrating their legacy knowledge base to knowledge-graph-based systems. Further, mapping to scalable, high-level ontology, such as the Function-Behaviour-Structure (FBS) ontology (Gero 1990), can support functional reasoning. Most research and development on automated knowledge extraction focuses on using extensive unstructured data sets, as they provide valuable contextual information. In an ideal scenario, enterprises and knowledge engineers would leverage these large unstructured data sets to create knowledge graphs (KG) that represent and organize their enterprise's product information. However, in practice, some of the most useful information is available in a structured data format, such as specification sheets and product catalogues. Such structured data sets do not follow a standard ontology, do not provide an explicit mapping of relationships between the entities, and provide no contextual information. Therefore, there is a need for techniques and methods for automated extraction and creation of ontology-based KG from structured legacy data that lack contextual information and ontological congruence. To address this gap, the following key challenges must be resolved:

- Extracting knowledge in terms of the target ontology-specific concepts and terminologies, which differ from the concepts and terminologies used in the source legacy structured data lacking contextual information. That is, extracting knowledge when there is no ontological congruence between the target KG and the source data set. For instance, the target KG in this research, the FBS ontology, requires descriptions and mappings in terms of the Function (F), Behaviour (B) and Structure (S) concepts and terminologies. Whereas, the source structured data set, such as the product categories, is described in terms of parts, specifications, and terminologies that are appropriate and known to vendors and customers but have no explicit mapping to the FBS concepts.
- The challenge in extracting targeted, ontology-specific KG from legacy structured data is compounded because legacy structured data does not explicitly map relationships between the entities and terminologies listed within the catalogue or specification sheet, nor does it provide any additional contextual information. That is, the challenge of dealing with the lack of ontological congruence is compounded by the lack of additional contextual information to build on.

- Automatically explicating knowledge in terms of target ontology-specific concepts when there is inadequate contextual information requires greater semantic clarity about the target concepts. In this case, the concepts of F, B, and S must be clearly defined because there is a limited contextual pattern to learn from the contextual data. The definitions and semantics of the F, B and S concepts need articulation and disambiguation, if any.

Accordingly, the following are the two key objectives and contributions of this research:

- The articulation and disambiguation of the FBS concepts to enable the target FBS-ontology-based KG to be created from a data set lacking contextual information.
- The development of a method and workflow for automated extraction and creation of the targeted FBS-ontology-based KG from a legacy structured dataset lacking contextual information.

The current work utilises specification sheets of product components to demonstrate the workflow's efficacy. The developed workflow was evaluated for its accuracy, generalizability, and applicability.

The following section provides an overview of ontologies relevant to the problem context, especially detailing FBS, formulation of guidelines and rules for semantic reasoning around FBS, and the Rule-based FBS-KG framework. The later sections discuss the workflow, its evaluation, and implications for research and practice.

# Background

The literature emphasizes the necessity of advanced methods for intelligent retrieval of design knowledge in the context of machine part design (Ning et al. 2024). Automated knowledge extraction and reuse have been used in different design and manufacturing activities. For instance, Helgoson and Kalhori (2012) employed CAD and CAM documentation for knowledge extraction. Shen et al. (2020)

focussed on representation for process planning. Others have concentrated on automating knowledge extraction from text to KG (Kertkeidkachorn and Ichise 2018), extracting knowledge from common domains (Zhang et al. 2017), developing a tool maintenance planning system (Wan et al. 2019), addressing process planning concerns (Šormaz and Sarkar 2019), and mining implicit information to update KG in real time (Zhou et al. 2021), characterizing high dimensional vibration characteristics (Qiu et al. 2024), and process plan generation (Zhou et al. 2022). A limitation of intelligent design knowledge retrieval methods is their reliance on the completeness and quality of the underlying data, which can hinder their effectiveness in novel or poorly represented scenarios.

Different approaches have been adopted using ontologies for automated design knowledge extraction/acquisition and representation. For example, Qin et al. (2017) developed a framework based on FBS for capturing and representing complex design knowledge to facilitate reuse in future design processes. Guo et al. (2022) developed a knowledge classification and extraction framework based on the Function Behaviour State ontology. However, these studies do not address the difficulties associated with source datasets being limited to structured data lacking contextual information.

Other approaches to integrating knowledge relevant to design tasks include capturing and tracing design knowledge during the design process (Tang et al. 2010), material selection (Zhang et al. 2017), supporting knowledge sharing and integration for design evaluation (Liu et al. 2019), and extracting technical and design information from patent-related documents (Huang et al. 2023). Zhang et al. (2021) proposed a framework for a feedback-driven push mechanism aimed at facilitating the exploration of eco-design-related knowledge essential for the design process. In a separate study, Wu et al. (2019) introduced an ontology-based approach for process knowledge sharing and recommendation, which was implemented in the context of mechanical parts design.

The studies conducted by Fantoni et al. (2013) and Huang et al. (2023) hold notable significance in ongoing discussions, as they revolve around the extraction and representation of F, B, and S from patent data and their practical implications. Both the works (Fantoni et al. 2013 and Huang et al. 2023) focus on patent information and utilise NLP for F, B, and S classification. Fantoni et al. (2013), based on a comprehensive review of the definitions of F, B, and S, presented a set of procedures and classification stages, thereby emphasizing the challenges associated with automating the extraction process from unstructured data and their practical implications. The dataset for these studies included unstructured data, which provides contextual information. This reliance on unstructured data restricts the integration of diverse data sources.

Similarly, multiple studies have used ontology-based approaches to support design generation. The Ideas Inspire framework (Chakrabarti et al. 2006) employed the FBS and SAPPHIRE models to index biological information and utilise them in design inspirations. Chen et al. (2012) presented a system for generating principle solutions from multidisciplinary knowledge based on a required function. Qin et al. (2017) proposed an ontological framework for identifying task definitions and reasoning. Related research includes recommender systems for design ideation (Dong et al. 2020; Hao et al. 2021; Liu et al. 2022), user knowledge-based conceptual design environment (Feng et al. 2011), and automated CAD model generation for mechanical products (Long et al. 2021). These investigations demonstrate the advancements in knowledge-based design generation by emulating the knowledge synthesis process. However, the main requirement in the above works is the reliance of predefined design knowledge.

## Overview of the FBS ontology

Gero (1990) introduced the FBS ontology as a design representation and reasoning model. They described function as requirements, behaviour as means to achieve the function through the structure and structure as elements of the artefact (Gero 1990). F-B-S ontology describes all designed things, or artefacts, irrespective of the specific discipline of designing to its fundamental concepts – Function (F), Behaviour (B) and Structure (S) (Gero and Kannengiesser 2004) as well as the critical aspect of the FBS ontology is its ability to provide abstraction at a more fundamental level while ignoring the non-essential detail. The FBS ontology is specifically designed to provide a detailed representation of the functional, behavioural, and structural aspects of entities within a domain. Thus, the FBS ontology theoretically allows a comprehensive representation of the knowledge in any domain. Gero's (1990) FBS framework, also called the design prototype, explains the end-to-end iterative design reasoning process using the FBS ontology, making FBS ontology a good candidate for automating the design generation process.

However, there have been several criticisms of the FBS framework regarding the ambiguity in the definitions and conceptual clarity of the key concepts, especially behaviour. Behaviour is assumed to manifest in structures, but it could also result from the combination of structure and environment (Vermaas and Dorst 2007). Vermaas and Dorst (2007) raise concerns over the descriptive and prescriptive aims of the FBS ontology. While the descriptive aim describes the human design process, the prescriptive aim can be used to automate design. If it is prescriptive, it should be useful and improve design. Vermaas and Dorst (2007) argue that the ambiguity in the FBS terminologies is a bottleneck in effectively utilising the FBS ontology as a prescriptive framework for automated design reasoning.

Umeda et al. (1990) argued that the ambiguity in the F-B-S definitions exists because they are generic terms used in different domains and serving different purposes. Vermaas and Dorst (2007) argue that the steps in the FBS model represent the trial-and-error method in design and need not follow the sequence of steps in the model. For instance, due to implicit knowledge, the functions could be directly translated into a structure without undergoing the FBS formulation step (Vermaas and Dorst 2007).

Despite these limitations and concerns, the FBS ontology is desirable in design reasoning because of its flexibility and generalizability. FBS can be applied at different decomposition levels of a system's hierarchy in a recursive approach. FBS can be mapped to other ontologies. Since the FBS framework has only three key concepts, it is cognitively more straightforward to apply. Most importantly, the FBS framework has been shown to explain the entire design process in iterative and recursive algorithmic steps (Gero 1990).

However, given the existing ambiguity in FBS definitions, the first step in the research is to develop a set of rules to reduce the ambiguity in classifying the information into F, B, and S variables.

## Available dataset and problem context

The available dataset and specific industrial case study for this research pertain to an electrical equipment supplier specialising in electrical, electronic, and optical connection, transmission, and networking, as well as manufacturing, mechatronics, and software creation. The case chosen was that of an electrical connector that consisted of four components- inserts, hoods, contacts, and housing.

# Research steps for developing the FBS classification guidelines

The F, B, and S classification guidelines were developed following the steps summarised in Figure 1.

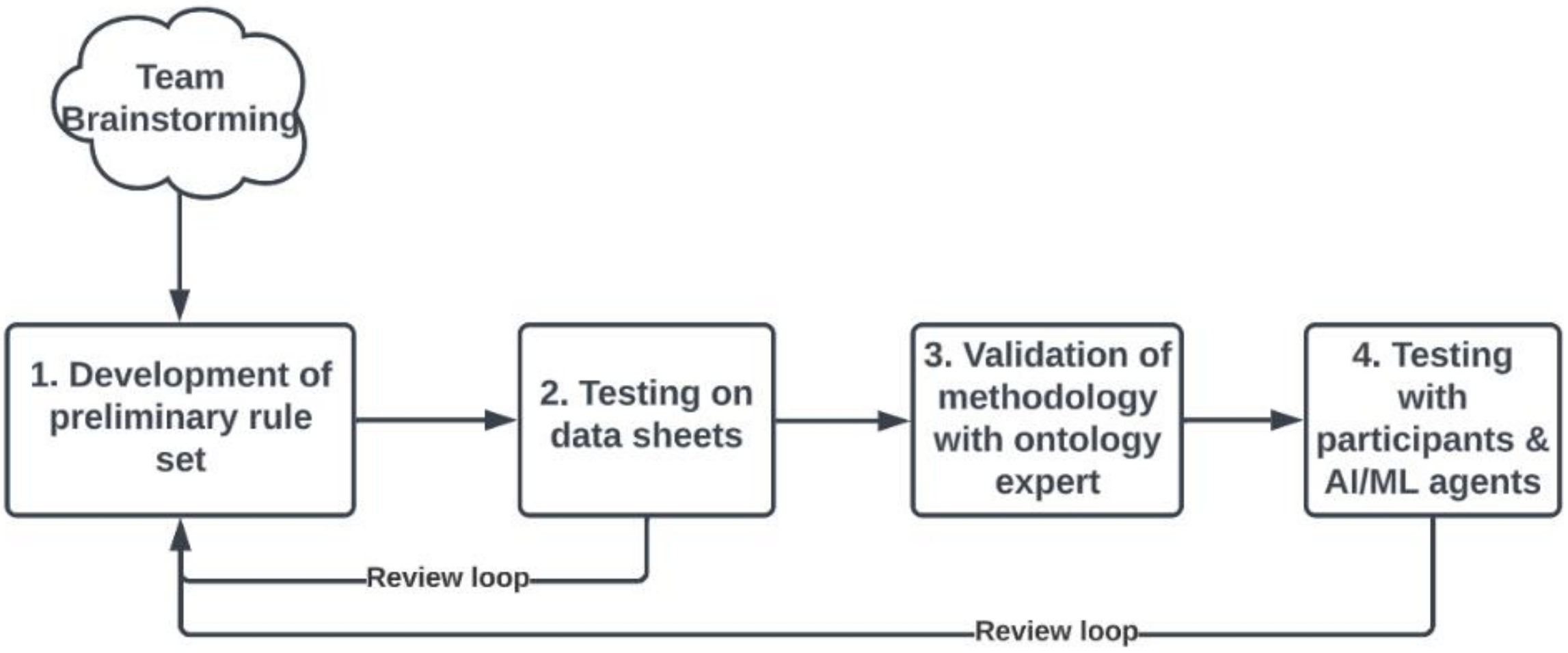

Figure 1: Development of FBS guidelines

To develop a set of guidelines for classifying an artefact into F, B, and S, multiple sessions were conducted with the problem domain experts to understand the components and to create a shared understanding of the FBS ontology. Multiple team brainstorming sessions and focus group discussions were conducted to discuss and refine the definitions of F, B, and S using examples. As the definitions started to be stable, the focus of the brainstorming and discussions shifted to determining the rules that could be used to extract the variables from the dataset.

Following the engagement with the problem domain experts, experiments with three experts in FBS ontology were conducted to review and assess the ease of applicability and accuracy of the developed rules. Each FBS expert was provided with all the extraction rules, examples of variables, and the datasheet for one of the components– contacts. In addition, an FBS mapping table and a Feedback form were provided. The rule sets were iteratively refined based on the feedback obtained from each expert. The accuracy percentage for the first, second, and third respondents were 54%, 93%, and 87%, respectively.

Following the sessions with the FBS ontology experts, a large-language model-based agent, the ChatGPT, was further used to test the rules. ChatGPT was passed a brief context about the project, followed by the rules. The agent was then provided multiple key-value pairs from the data sheet and was tasked to classify them based on the rules. The results showed about 78% accuracy of classification. The rules were further refined and retested with ChatGPT based on the above. The output exhibited 100% accuracy.

Table 1 summarizes the definitions and rules derived from the above-described steps.

Table 1: Rules for FBS classification

| Variable | Rule | Definition | Examples |
|---|---|---|---|
| Function | 1. Any intended purpose, application, or adherence to standards can be considered its Function. | The intended use, purpose, or goal of the artifact | • The main function of a pencil is to leave a mark on the paper.<br>• The main function of a chair is to provide seating. |
| Behaviour | 1. Any property with a unit of measurement can be considered a behaviour within the FBS ontology. | The characteristics, attributes, or properties emanating from the artifact resulting in or allows the intended function | • The size of a pencil can be measured in inches or centimeters and can be considered as a behaviour of the pencil.<br>• The weight of a chair can be measured in pounds or kilograms and can be considered as a behaviour of the chair. |
| | 2. The characteristics of an entity in operation, such as its working behaviour, can be considered a behaviour. | | • The way a pencil transfers its lead onto a sheet of paper is its behaviour.<br>• The way a chair supports the weight of a person sitting on it is a behaviour of the chair. |
| | 3. Properties that exhibit classification, types, or sub-classes should be deemed a Behaviour. | | • Different types of windows, like sliding or hinged windows, can be seen as a Behaviour. |
| Structure | 1. Any entity, sub-entity, or configurational aspect of an entity can be considered as a Structure within the FBS ontology. | An element/component or the configuration of elements/ components of the artifact that hold the derivable behaviour | • The components of a chair, such as legs, arms, seat, etc, are its structures. The chair itself is a structure.<br>• The material of a chair can be considered a Structure. The material determines what behaviours one can expect from the chair.<br>• The configuration of elements is also a structure. For instance, diamond and graphite |
| | | | exhibit very different behaviours despite being made of carbon atoms because of their different structure or configuration. |

The refinement in definitions, though subtle, improved conceptual and causal consistency. For instance, the reformulated guidelines led to a reclassification of geometry as a behaviour, contrasting with prior studies where it was categorised as a structure. This shift stemmed from recognising that an artefact's geometry is contingent upon the assumption of its rigidity, a behaviour. According to the revised rules, anything quantifiable is deemed a behaviour.

# Rule-based FBS-KG Framework

The developed Rule-based FBS-KG Framework is shown in Figure 2. The core of the framework is the workflow for population and retrieval of the FBS-based KG. The inputs to the architecture are the specification sheets of the components, the FBS-based KG structure, and the FBS classification rules.

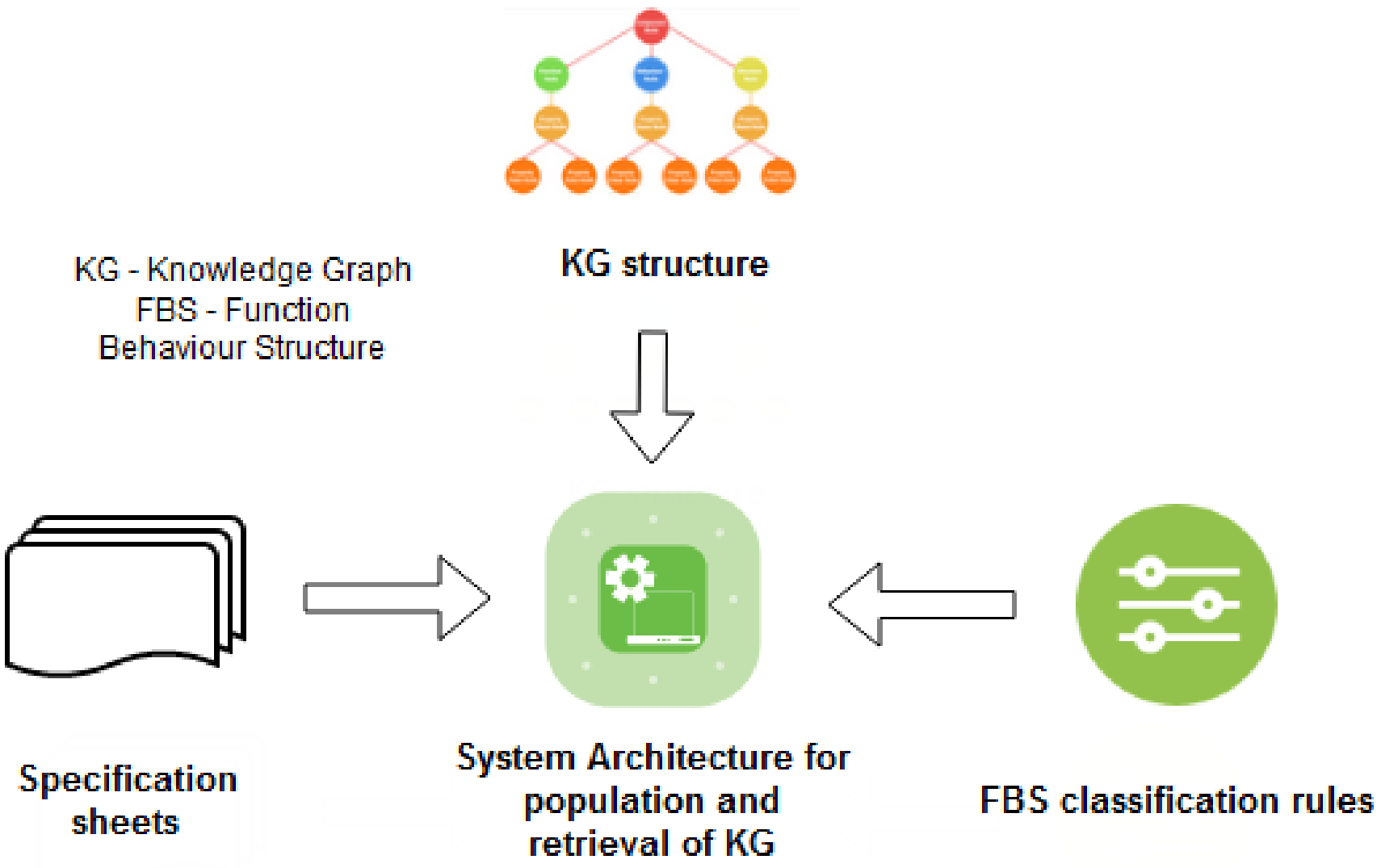

Figure 2: Rule-based FBS-KG Framework

The KG serves as a rich representation of product information, organized into categories as defined by the ontology, allowing for efficient traversal. Neo4j graph database is the platform chosen for the sample implementation in this work.

The graph can be construed as being made up of disjoint trees of nodes, wherein each tree represents a single component of a more extensive configuration, for example, Contact. Each tree has a depth of four distinct nodes, each related to the other using directed edges. Each node is an instance of one of four types, as shown in Figure *3*.

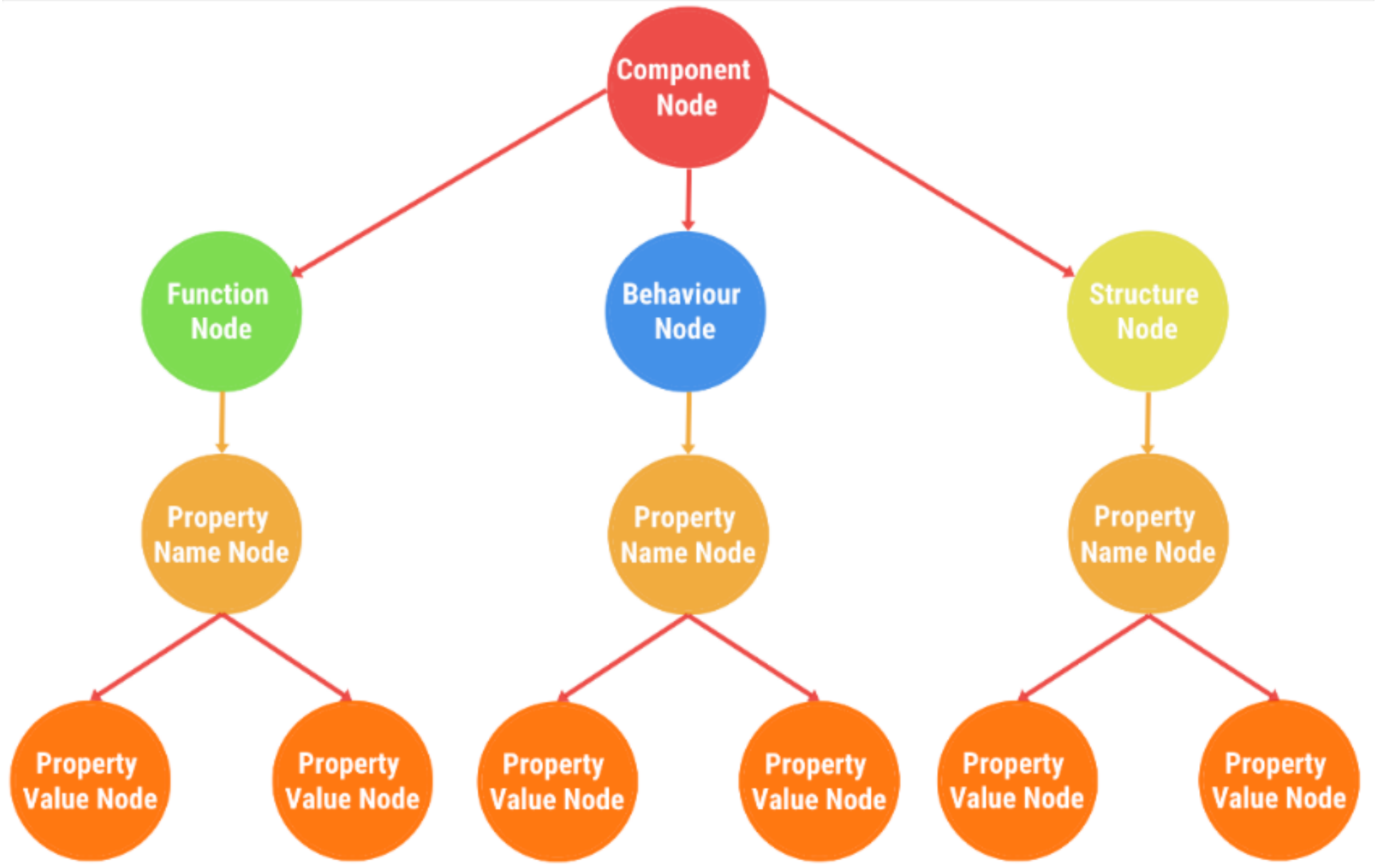

Figure 3: The disjoint FBS KG Structure

The properties of the nodes are summarized in Table 2.

Table 2: Nodes in the Neo4j KG

| **Type / Description** | **Component Name Node (root node)** | **FBS Class Node (internal node)** | **Property Name Node (internal node)** | **Property Value Node (leaf node)** |
|---|---|---|---|---|
| Definition | Denotes one category of products | Denotes one of the 3 classes that a property may belong to | Denotes one kind of property of its associated component | Denotes one value of the concerned property. Is associated with one model instance of a component |
| Label | "Component" | "Function", "Behaviour" or "Structure" | "PropertyName" | "PropertyValue" |
| Attribute(s) | name - name of component. For example, "Contact" | - | name - Name of property. For example, "Rated Voltage" | 1. unit - Unit of measurement, where applicable. For example, "A"<br><br>2. value - Value of measurement, for fixed value properties. For example, "10.0"<br><br>3. lower_limit_value - Lower limit of accepted values of a range of applicable values. For example, "0.0"<br><br>4. upper_limit_value - Upper limit of accepted values of a range of applicable values. For example, "100.0" |

Each individual node has a unique `<id>` associated with it. If two or more models of a single component have the same value of a given property, they share a common property value node. The various aspects of the edge are summarised in Table *3*.

Table 3: Edges in KG

| Type / Description | Component to FBS | FBS to Property Name | Property Name to Property Value |
|---|---|---|---|
| Definition | (Component, FBS Class) | (FBS Class, Property Name) | (Property Name, Property Value) |
| Label | "HAS" | "YIELDS" | "HAS" |
| Attribute(s) | - | - | 1. model_name - the name of the model with whom this property value is associated<br><br>2. part_number - the part number of the model with whom this property value is associated |

The following sub-section introduces and expands upon the design of the pipelines used for population and querying of the KG, with a detailed description of each step.

***Workflow Development***

The system architecture for the workflow is shown in Figure 4. The developed workflow aims to achieve two purposes:

- To extract information from specification sheets and populate the FBS based KG.
- To retrieve configurations based on user-provided queries.

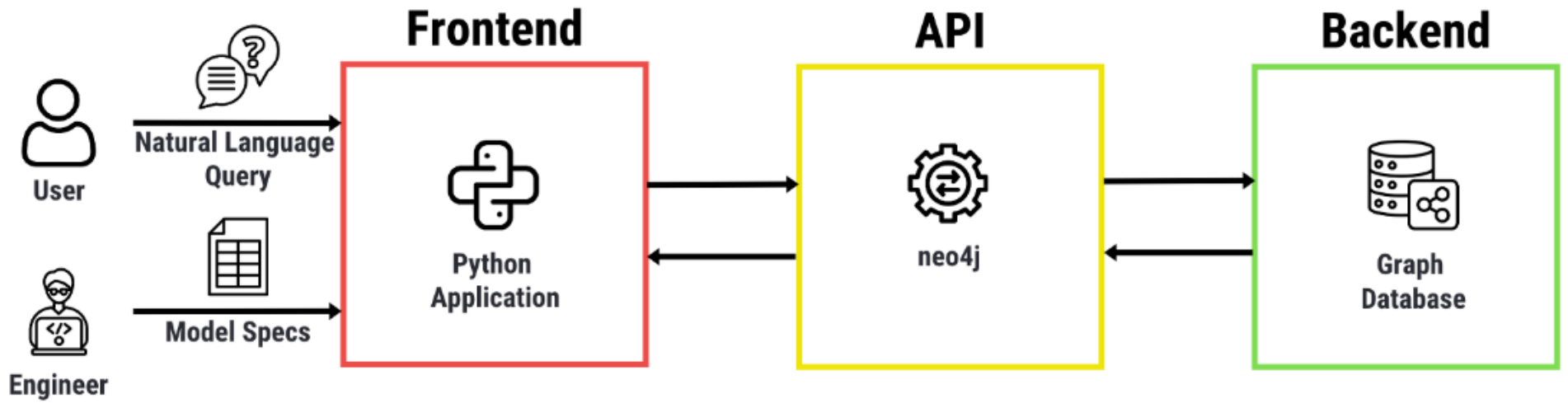

Figure 4: System Architecture of the workflow

The overall architecture comprises the frontend, application programming interface (API), and backend. The front-end extracts data and processes it into a form that can be represented in a KG. It then generates queries the API passes to run on the graph database that forms the backend. These queries function to either populate the database using the specification data sheets uploaded by engineers or retrieve insights requested by the user using natural language.

***Knowledge Graph Retrieval System Design and Development***

The process of creation and utilisation of a KG typically follows a sequence of three steps: extraction of usable data and its relationships, representation of information into knowledge, and derivation of actionable insights. The system for retrieving the identifying information of the target component models in the current work comprises two constituent pipelines:

- KG population pipeline
- KG querying pipeline

The first pipeline is concerned with populating the KG with all models of every component in the problem domain. The second handles the generation and execution of cypher queries that query the KG hosted on the Neo4j instance to retrieve the identifying information of all target models. Figure 5 shows the mapping between the sequence of steps and the pipelines associated with the KG.

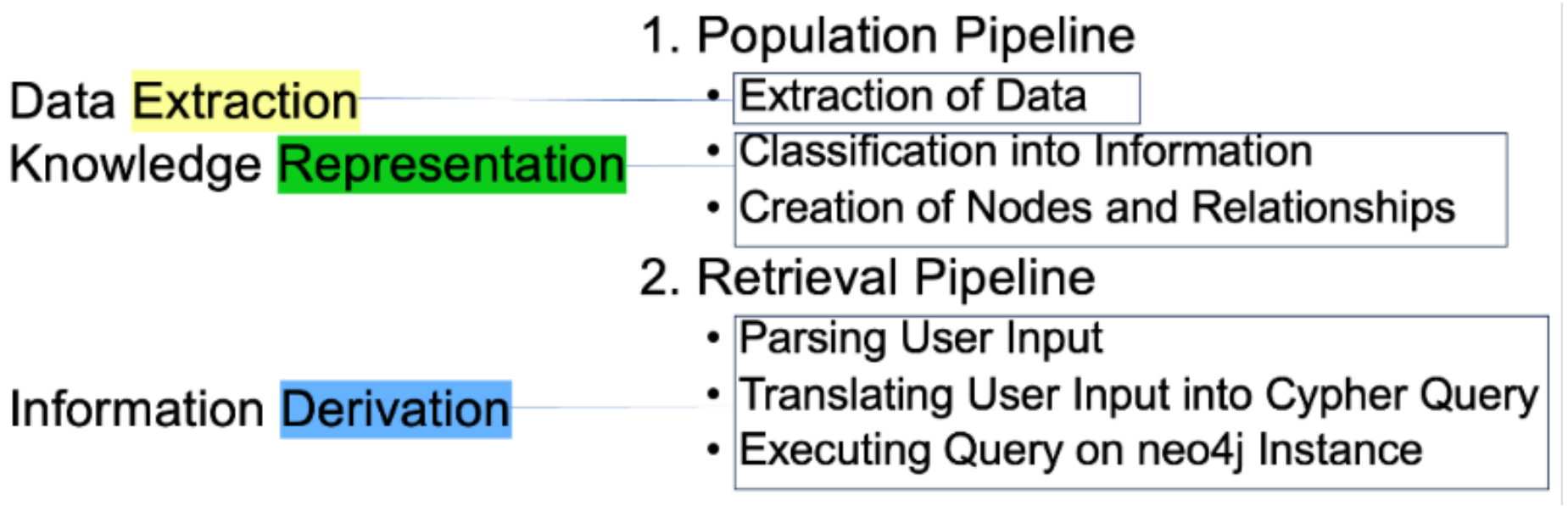

Figure 5: Mapping between sequence and pipelines

Figure 6 illustrates the flow of information through the various components that comprise the system, starting with the model specification data sheets and ending with product information.

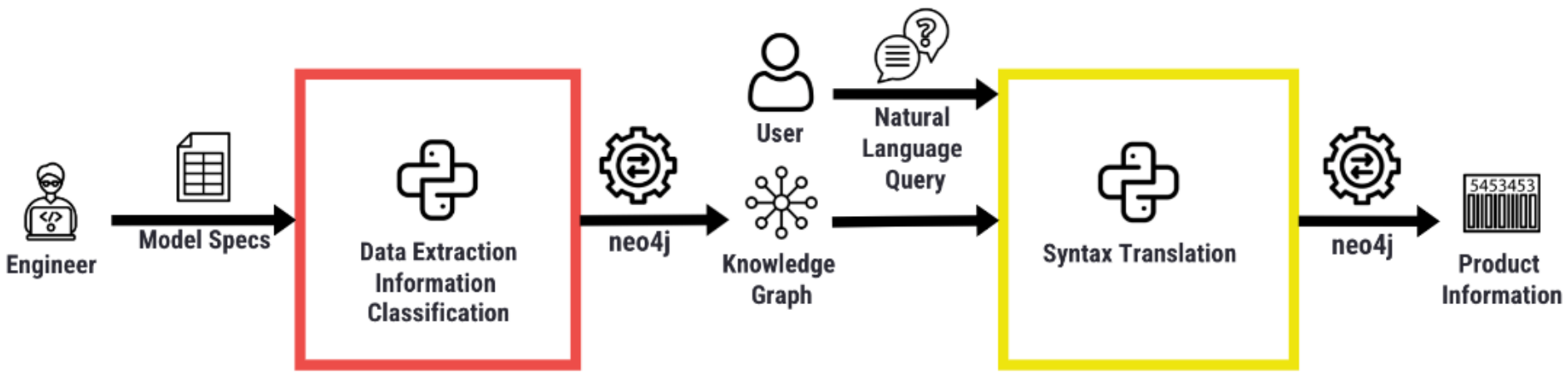

Figure 6: Data flow

***KG Population Pipeline***

As shown in Figure 7, the KG population pipeline starts with a product (model of a component) data sheet. It ends with a new or updated graph, repeated for every new product to be included in the knowledge base.

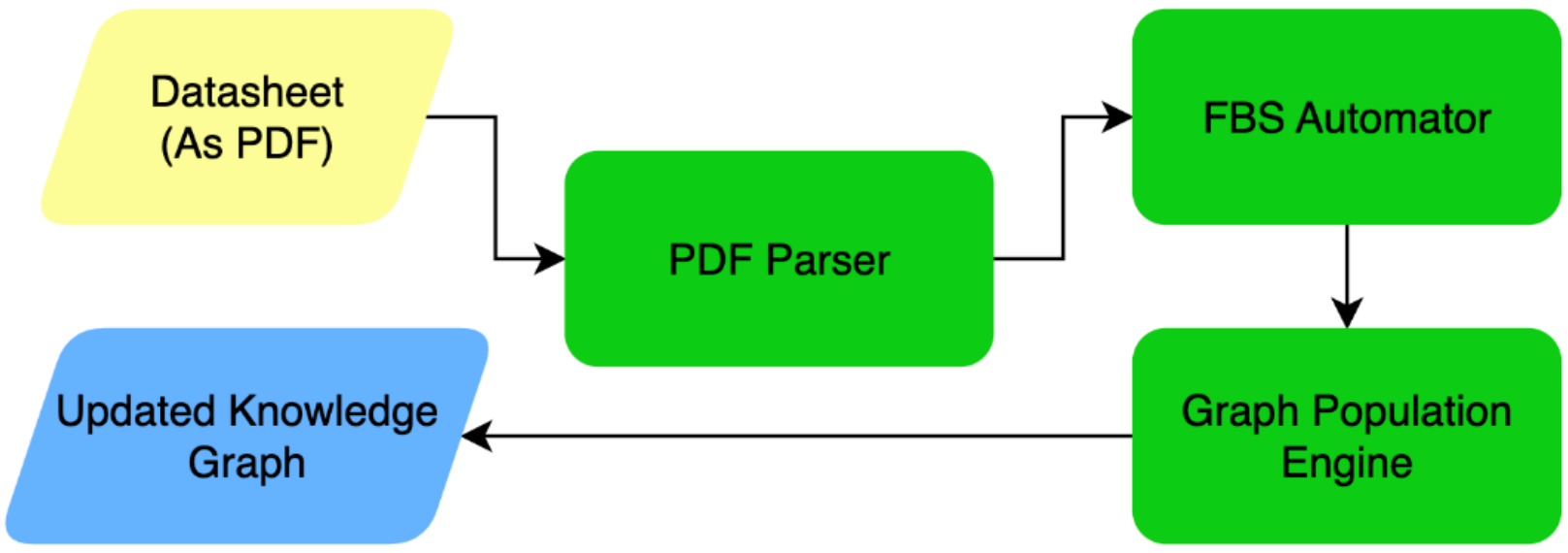

Figure 7: KG population pipeline

**Datasheet (As PDF):** The input to the pipeline is a set of product data sheets, represented in PDF form, that detail the specifications of a particular component model. Each datasheet consists of tables under several headings, each denoting a type of specification relating to the material properties, technical characteristics, identification, conformity to standards, etc. The tables are composed of two columns, indicating the name and value of each property that falls under the respective heading.

Each row of the table represents a single property. Notably, each property may have several associated values, such as a list of specifications under the conformity to standards section. Every value occupies its own sub-row within each property row, an important aspect for the next step of the pipeline.

**PDF Parser:** The input to this step is a product datasheet in PDF form consisting of model specifications. A Camelot-py library is used to parse the PDF and render its contents in a pandas DataFrame called specs_df with two columns - property name and property value. The library uses image processing techniques to read each line of text in the PDF, splitting them into columns wherever the outline of a boundary is detected.

There are cases in which a property may have multiple values associated with it, each occupying its sub-row, which in most cases had no associated property name. As a result, these values were each considered part of separate, non-existent properties during parsing. To resolve this issue, an algorithm was developed to associate such 'orphan' property values with their property names as part of the same property.

Following this, irrelevant text read by the parser, including pagination footers, is detected using a regular expression and removed from the dictionary `result.` Finally, a list of key-value pairs is constructed from the `result`, which is then converted back into a pandas `DataFrame` and assigned to `specs_df`. The output of this step is the pandas `DataFrame specs_df` which contains all property names and associated property values of the model.

**FBS Automator – Operation:** The pandas `DataFrame specs_df` is input for this processing step. Each property in `specs_df` is classified into one of four categories: F, B, S, or Unknown.

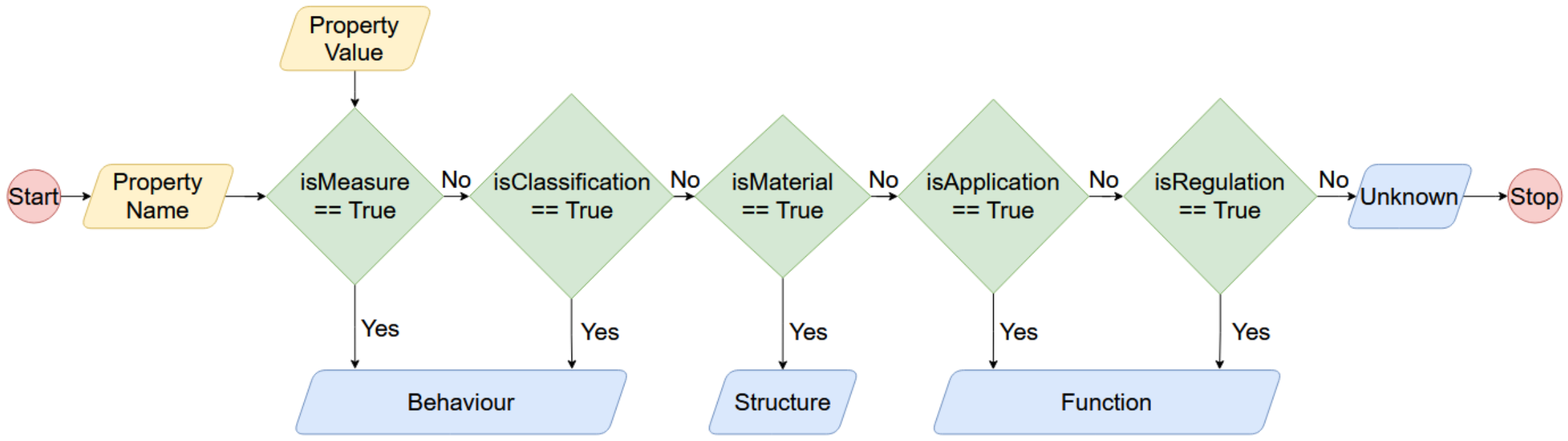

Figure 8: FBS Automator

The classification is performed using various NLP techniques, as detailed below. In the current work, the measure and classification property was classified as behaviour, the material property was classified as structures, and the application and regulation properties were classified as functions. This is based on the FBS guidelines and rules derived as discussed earlier. The details of the logic used in the categorisation are provided in Appendix I.

**Graph Population Engine:** This is the primary step in the KG population pipeline. The logic used in this step is shown in the flowchart in Figure 9. The corresponding algorithm is shown in Figure 10. The logic outlines the mechanism by which:

- A new graph is created from scratch if it does not already exist, and
- An existing graph is updated with information about a new model of the same component if the graph for that component already exists, taking care not to duplicate information already present in the property name and value nodes of the previous model(s).

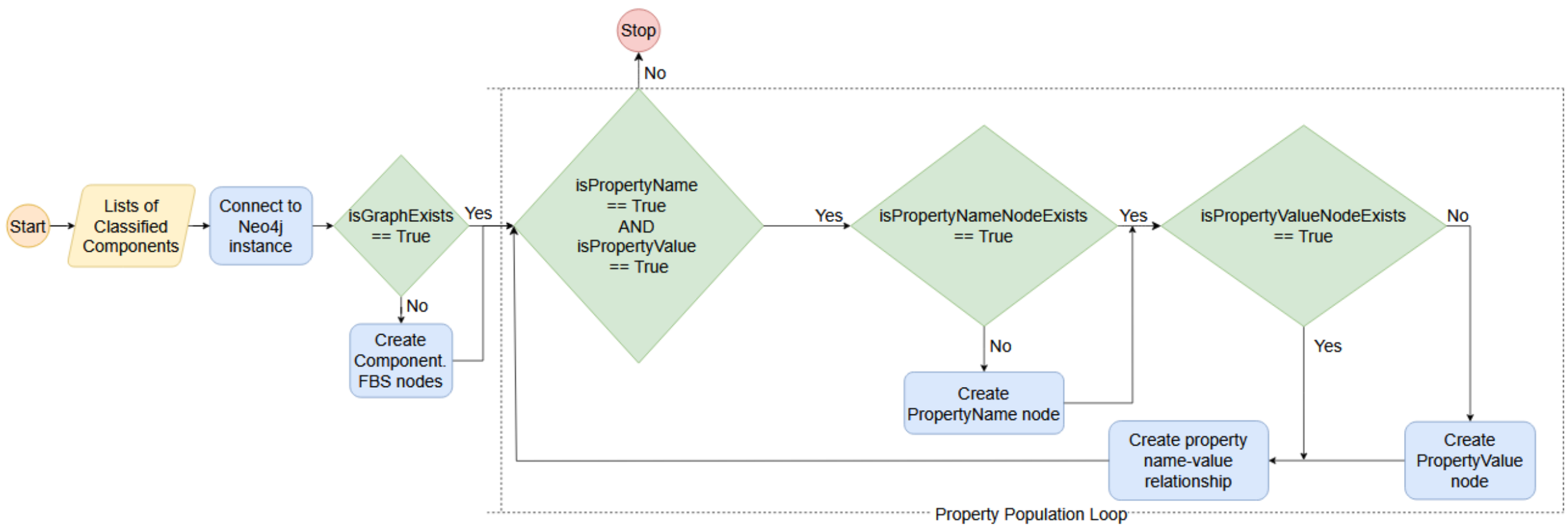

Figure 9: Graph Creation Engine

- Input
  - Dictionary of lists, one for each facet of FBS
- Process
  - Employ the `py2neo` library to abstract some functionality
  - Connect to `Neo4j` instance
  - If graph does not exist,
    - Create component and FBS facet nodes
    - Iterate through each facet list
      - Populate property name and value nodes from scratch
      - Create a relationship between the above nodes, and assign appropriate part and model numbers as attributes
  - If graph already exists,
    - Fetch component and FBS facet nodes
    - Iterate through each facet list
      - If property name does not exist, create property name and value nodes from scratch
      - If property name exists, check if values match
        - If they match, create new relationship to existing value node and assign appropriate part and model numbers as attributes
        - If they don't match, create new value node and a relationship to it, assigning appropriate part and model numbers as attributes
- Output
  - Created / Updated Knowledge Graph

Figure 10: Algorithm for FBS-based KG population

This step involves the population of the KG hosted on the Neo4j instance with all previously classified properties of the current model under consideration.

**Updated KG:** The final output of the KG population pipeline is a new or updated KG, hosted on a Neo4j AuraDB instance in the cloud.

***Knowledge Graph Retrieval Pipeline***

Figure shows steps by a natural language query that transforms it into a Cypher query format that can be run on the Neo4j instance to retrieve the part number(s) of the target product(s) from the KG. The details of each of the steps are described below.

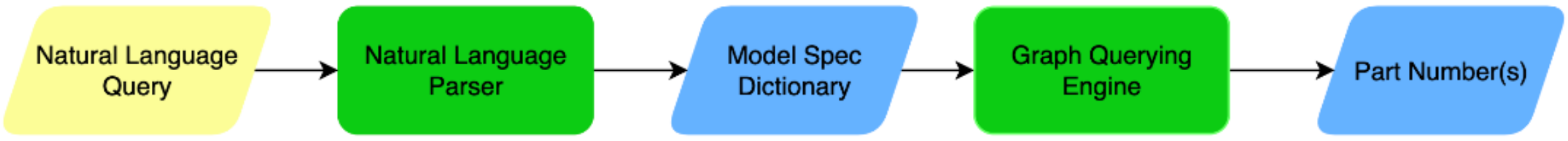

Figure 11: KG retrieval pipeline

**Natural Language Query:** The input to the pipeline is a natural language query in plain English, following a standard format. The example query used to showcase the operation of the system is as follows :- "`Give me a crimp contact that conforms to these specifications :- Material (contacts) equal to Copper alloy, Surface (contacts) equal to Silver plated, Conductor cross-section-2 between 22 and 26, Gender = Female, Manufacturing process = Turned contacts, Operating Current lesser than or equal to 10, Contact resistance ≤3, Mating cycles greater than or equal to 500, Stripping length = 8`". Thus, the query specifies the target model type and the set of identifying properties (names and values) to be determined from the graph.

**Natural Language Parser:** This processing step takes the natural language query as its input. Firstly, a regular expression is employed to detect and isolate the model_type. Secondly, another regular expression is used to extract groups of property_name, relational operators, and property_value from the text. Each group contains all the information necessary to match a single property in the target product(s). A Python dictionary named model_spec is created, whose keys hold property names and values, which are sub-dictionaries containing values to match and the concomitant relational operator to be used for comparison.

The sub-dictionaries are classified into three types depending on the nature of the matched values: single-valued, single-limit ranged, and dual-limit ranged. The relational operator is used to select the bound (lower or upper), in case of single limit range values. The output of this step is the Python dictionary `model_spec.`

**Model Spec Dictionary:** The Python dictionary model_spec will be parsed to yield a Neo4j cypher query in the following steps. It contains all property names, value pairs that should be present in the target product(s).

**Graph Querying Engine:** The Python dictionary model_spec is input to this operation and consists of two sub-operations: Generating cypher query and executing cypher query. The algorithm for generating Cypher Query is shown in Figure 12.

- Define variable `query` of type string
- For each property name,
    - Add match condition for property name
    - Add match conditions for each type of property value
- Add conditions to check if `model_name` associated with each `HAS` property name to value relationship is the same
- Add condition to check if `model_type` is a substring of `model_name`
- Return `part_number` associated with `HAS`

Figure 12: Algorithm for Cypher Query

Algorithm 2 lays out the steps followed in populating the cypher query with three sets of conditions to be met. The first constitutes a set of match conditions, one for each property, considering the different types of property values (single-valued, single-limit, and dual-limit ranges). The second part uses the `model_name` attribute of the relationship `HAS` between a property name and its associated value to check if all matched properties belong to the same model. The third part checks if the matching `model_name` contains the `model_type` as a substring. Finally, a statement to return the values of `part_number` for all products that match all the conditions is included.

Execute cypher query: The `query` thus generated is then run on the Neo4j AuraDB instance hosted on the cloud. A list of `part_number` values of products that match all conditions specified in the query is returned as the output of the graph querying engine.

**Part Number(s) - Final Output:** A list of `part_number` values of all matched models of `Component`

***Workflow Testing***

The data sheet of a component called the ‘male contact’ was used to test the workflow. This section only provides the instances of the outputs from the workflow. Figure 13 shows the specification sheet for the component.

Han TC100 male contact 35 mm² (Han K6/6)

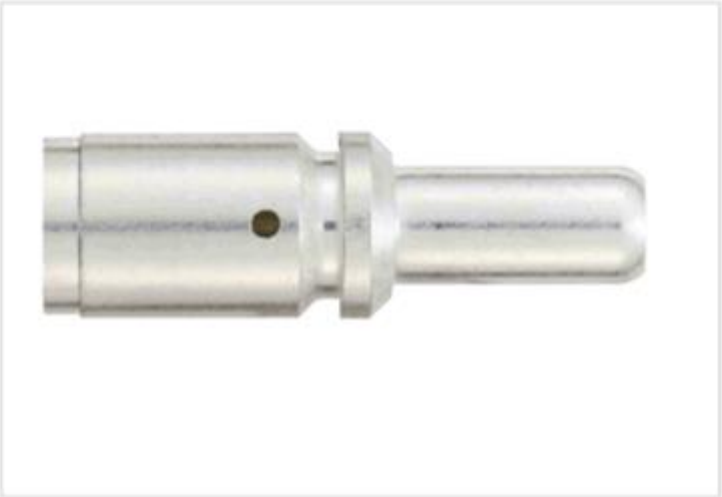

Image is for illustration purposes only. Please refer to product description.

Identification

| | |
|---|---|
| Category | Contacts |
| Series | TC 100 |
| Type of contact | Crimp contact |

Version

| | |
|---|---|
| Gender | Male |
| Manufacturing process | Turned contacts |
| Details | Only for use in the Han® K 6/6 |

Technical characteristics

| | |
|---|---|
| Conductor cross-section | 35 mm² |
| Operating current | ≤100 A |
| Contact resistance | ≤0.3 mΩ |
| Stripping length | 16 mm |
| Limiting temperature | -40 ... +125 °C |
| Mating cycles | ≥500 |

Material properties

| | |
|---|---|
| Material (contacts) | Copper alloy |
| Surface (contacts) | Silver plated |
| RoHS | compliant with exemption |
| RoHS exemptions | 6(c): Copper alloy containing up to 4 % lead by weight |
| ELV status | compliant with exemption |

Figure 13: Example data sheet

KG population output is shown in Figure 14. The KG classification, population, and retrieval snapshots are provided in Figures 15 to 17.

```
Parsing pdf...
PDF Parsed.

** FBS Automator **

Classifying properties...
** Behaviour classifier **
Entities []
☑ GENDER is a Behaviour (classification-oriented)
Similar to keyword: gender
Entities []
☑ MANUFACTURING PROCESS is a Behaviour (classification-oriented)
Similar to keyword: classification
Entities [('cross-section', 'MEASURABLE QUANTITY')]
☑ CONDUCTOR CROSS-SECTION is a Behaviour (unit of measure)
Entities [('cross-section-2', 'MEASURABLE QUANTITY')]
☑ CONDUCTOR CROSS-SECTION-2 is a Behaviour (unit of measure)
Entities [('current', 'MEASURABLE QUANTITY')]
☑ OPERATING CURRENT is a Behaviour (unit of measure)
Entities [('resistance', 'MEASURABLE QUANTITY')]
☑ CONTACT RESISTANCE is a Behaviour (unit of measure)
Entities [('length', 'MEASURABLE QUANTITY')]
☑ STRIPPING LENGTH is a Behaviour (unit of measure)
Entities [('cycles', 'MEASURABLE QUANTITY')]
☑ MATING CYCLES is a Behaviour (unit of measure)
Entities [('Material', 'STRUCTURE')]
✕ MATERIAL (CONTACTS) is not a Behaviour
Entities [('Surface', 'STRUCTURE')]
✕ SURFACE (CONTACTS) is not a Behaviour
Entities []
✕ ROHS is not a Behaviour
Entities [('exemptions', 'MEASURABLE QUANTITY')]
☑ ROHS EXEMPTIONS is a Behaviour (classification-oriented)
Similar to keyword: classification
Entities []
```

Figure 13: KG population output (page 1)

| Property Name | FBS Ontology Classification |
|---|---|
| California Proposition 65 substances | Behaviour |
| California Proposition 65 substances-2 | Behaviour |
| China RoHS | Behaviour |
| Conductor cross-section | Behaviour |
| Conductor cross-section-2 | Behaviour |
| Contact resistance | Behaviour |
| ELV status | Behaviour |
| Gender | Behaviour |
| Manufacturing process | Behaviour |
| Mating cycles | Behaviour |
| Net weight | Behaviour |
| Operating current | Behaviour |
| RoHS exemptions | Behaviour |
| Specifications | Behaviour |
| Stripping length | Behaviour |
| ECHA SCIP number | Function |
| Image is for illustration purposes only. Please refer to product description. | Function |
| REACH ANNEX XIV substances | Function |
| REACH Annex XVII substances | Function |
| REACH SVHC substances | Function |
| REACH SVHC substances-2 | Function |
| RoHS | Function |
| Technical characteristics | Function |
| conformity with applicable laws and directives, as well as for the electrical safety in the particular application. | Function |
| HARTING Stiftung & Co. KG \| Marienwerderstr. 3 \| 32339 Espelkamp \| Germany | Structure |
| Han D F Crimp Contact Ag AWG 26 - 22 | Structure |
| Han D F Crimp Contact Ag AWG 26 - 22-2 | Structure |
| Material (contacts) | Structure |
| Material properties | Structure |
| Material properties-2 | Structure |
| Surface (contacts) | Structure |

Figure 145: Summary of FBS classification

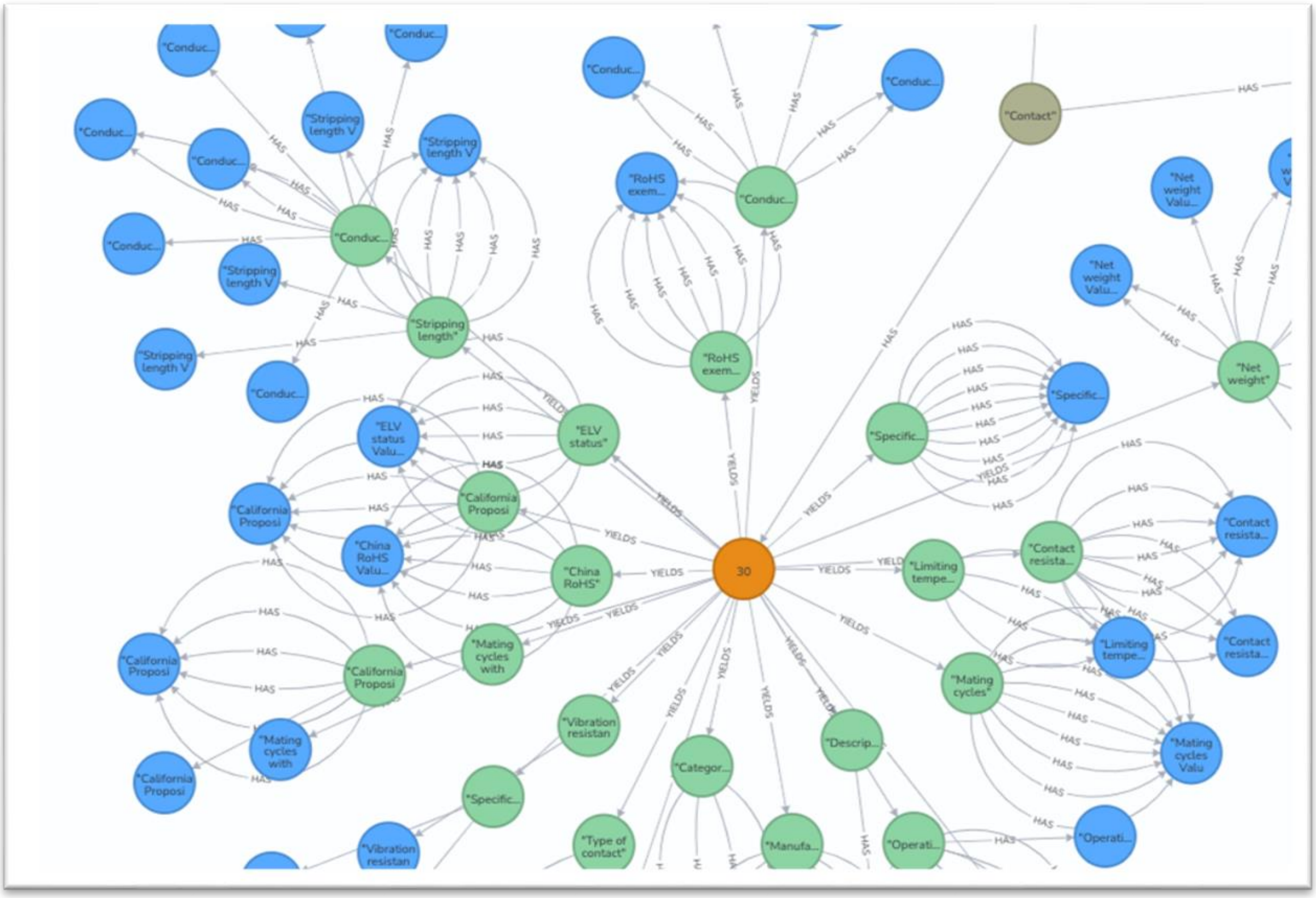

Figure 16: FBS based Knowledge Graph

```
MATCH (`Mating cycles`:PropertyName)-[has7:HAS]->(`Mating cycles_value`:PropertyValue)
WHERE toLower(toString(`Mating cycles`.name)) = 'mating cycles' AND toFloat(`Mating cycles_value`.upper_limit_value) >= 500 OR (`Mating cycles_value`.
lower_limit_value IS NOT NULL AND toFloat(`Mating cycles_value`.lower_limit_value) >= 500)

MATCH (`Stripping length`:PropertyName)-[has8:HAS]->(`Stripping length_value`:PropertyValue)
WHERE toLower(toString(`Stripping length`.name)) = 'stripping length' AND toFloat(`Stripping length_value`.value) = 8 OR (`Stripping length_value`.low
er_limit_value IS NOT NULL AND toFloat(`Stripping length_value`.lower_limit_value) >= 8) OR (`Stripping length_value`.upper_limit_value IS NOT NULL AN
D toFloat(`Stripping length_value`.upper_limit_value) <= 8)

MATCH (`Conductor cross-section-2`:PropertyName)-[has9:HAS]->(`Conductor cross-section-2_value`:PropertyValue)
WHERE toLower(toString(`Conductor cross-section-2`.name)) = 'conductor cross-section-2' AND (toFloat(`Conductor cross-section-2_value`.value) >= 22 OR
 (`Conductor cross-section-2_value`.lower_limit_value IS NOT NULL AND toFloat(`Conductor cross-section-2_value`.lower_limit_value) >= 22)) AND (toFloa
t(`Conductor cross-section-2_value`.value) <= 26 OR (`Conductor cross-section-2_value`.upper_limit_value IS NOT NULL AND toFloat(`Conductor cross-sect
ion-2_value`.upper_limit_value) <= 26))

WITH has1, has2, has3, has4, has5, has6, has7, has8, has9
WHERE has1.model_name = has2.model_name
AND has2.model_name = has3.model_name
AND has3.model_name = has4.model_name
AND has4.model_name = has5.model_name
AND has5.model_name = has6.model_name
AND has6.model_name = has7.model_name
AND has7.model_name = has8.model_name
AND has8.model_name = has9.model_name
AND toLower(has1.model_name) CONTAINS 'crimp contact'

RETURN DISTINCT has1.part_number

Running query on Neo4j instance...

Result:-
Part numbers of matching products :-
09 15 000 6204
```

Figure 17: KG querying output (page 2)

Owing to the paucity of space, detailed output results are not included.

# Workflow Evaluation

The framework was evaluated against two criteria, 1.

- The performance of the classification model for obtaining F, B, and S using confusion matrices.
- The perceived usefulness of the workflow to the end user, the industry partner. The evaluation of the model was carried out iteratively throughout the development of the workflow. This section summarizes the results obtained from the final evaluation.

***Evaluation of the classification model's performance:*** The classification model's performance was assessed using the Accuracy, Precision, Recall, and F1 score indices.

Table X: Scores from evaluation of the classification model’s performance

| | **Precision (%)** | **Recall (%)** | **F1 score (%)** | **Support** |
|---|---|---|---|---|
| **Function** | 73 | 100 | 85 | 11 |
| **Behaviour** | 57 | 57 | 57 | 7 |
| **Structure** | 78 | 54 | 64 | 13 |
| **Unknown** | 0 | 0 | 0 | 0 |
| **Accuracy** | | 71 | | 31 |
| **Macro average** | 69 | 70 | 68 | 31 |
| **Weighted average** | 72 | 71 | 70 | 31 |

Based on the observed accuracy and the averages, it can be inferred that the classifier model provides reasonable accuracy in classification for the partner company. The model performed better in classifying Functions than the other two variables. Behaviour classification had the lowest performance.

***Evaluation of the workflow's usefulness:*** The workflow's usefulness was evaluated at each step of the workflow development. A smaller R&D team was involved in biweekly meetings and evaluations in the initial stages. The final assessment was conducted by a larger team involving senior management and their knowledge management experts. The team found the workflow useful and applicable in their knowledge management practice.

# Discussion

A review of the existing ontological frameworks revealed the FBS ontology's advantages in ease of application, generalizability, and capturing the overall design-reasoning steps. To overcome the limitations of the FBS ontology in terms of ambiguity in its definitions, guidelines for FBS classification needed to be developed. While the rules proved reasonably effective in the current case of electrical configurator components, applying the rules to a different case might necessitate adjustments.

The focus of this work was to develop a method and workflow for automated extraction and creation of the targeted FBS-ontology-based KG from a legacy structured dataset. The evaluation of the framework gave indications regarding the effectiveness of the developed framework. The confusion matrix showed reasonable overall accuracy in classification but revealed differences in performances across F, B, and S variables. The model exhibited relatively accurate identification of the F variables with variations in the accuracy of classification of B and S. The results indicate the need for further refining the rules, particularly in classifying the B and S variables. A necessary investigation towards the comprehensiveness of the rules to accurately classify data of various formats needs to be conducted.

The investigation highlighted two crucial aspects in the discourse on applying FBS ontology. While the functional decomposition of the product into the three variables was achieved, further decomposition was not pursued in the current work. Additionally, it was observed that the structure of the KG contained only vertical linkages. Developing an approach to identify cross-linkages between behaviours is essential for enabling knowledge synthesis and reasoning-related tasks. This could enable a more accurate KG representation of a product from its documentation.

A crucial necessity within an ontological framework is the seamless applicability for users. The present work focuses on this aspect and suggests an approach that involves furnishing users with rules for the framework's application. With robust guidelines, it is anticipated that the FBS can provide a shared language for representing all forms of design knowledge, including design rationale, constraints, environment, and scientific principles. Further, it can be applied to other types of industrial processes to drive process innovations.

The results revealed that the Rule-based FBS-KG framework effectively automates the process of populating and retrieving information using FBS-based ontology from legacy data. This suggests that documented product information can be leveraged to construct a KG for formal representation. Given the role of KG as a recommender system, this marks the first step towards automating design generation and driving innovation within the FBS framework. In addition, it was observed that the specification sheets did not contain the design information in terms of first principles. However, it is also known that all design decisions are not based on first principles but could be based on intuitive relationships. Hence, the discernment of the mechanisms of design reasoning could be a topic of future investigation.

# Conclusion

The current work presents a method towards the automated FBS-based knowledge representation and retrieval framework from structured legacy dataset. Towards this end, a set of FBS guidelines, an FBS-based KG structure, and a workflow for populating and retrieving FBS-based KG were designed, developed, and evaluated. The framework was developed in collaboration with an industry partner and implemented in the case of an electrical configurator. The implementation results revealed that the framework showed reasonable effectiveness in using structured data legacy data for populating and retrieving information from FBS-based KG.

This framework provides the foundation for automated design generation using automated FBS reasoning. Future work will focus on enhancing the system's ability to generate design solutions. This would require further investigation through the development of the FBS reasoning-based automation system.

# Appendix-I

**`Behaviour` - based on property name**

A property is classified as type of `Behaviour` if it can be either

a. `Measure` or

b. `Classification`

**`Measure` - based on property name**

A sequence of conditions undertakes to determine if the property constitutes a measurable quantity or not. In parallel, checks are performed to discern whether the value of the property, should it prove to be a `Measure`, is single-valued or a range of values, with or without associated units of measurement, to enable different representations in property value nodes in the graph while populating them in the next step downstream. Regular expressions can be used to detect such patterns, but their presence is not conclusive proof of the property being a `Measure`. Consider, for instance, the following two properties, with ‘Assembly instructions’ and ‘Size’ names. The former has a value of 'At voltages of < 5 V and currents < 5 mA, gold-plated contacts are recommended’, and the latter, ‘1 A’.

If analysis were to be restricted to just the mentioned property values, the presence of a range of permitted values and an ostensible unit of measurement would merit a classification of `Measure`. This is, however, incorrect as the first represents a specific recommendation and the second a model number, not a measurement. To account for this, it was determined that analysis of the property name would yield valuable context regarding the property's status as being a measurable quantity or not.

To this end, an NER model trained on a small dataset of relevant property names was used to classify the property name attribute as either a `Measurable Quantity`, `Immeasurable Property` or `Model Number`. The first would capture the intended properties to be classified as a `Measure`, while the second and third capture values like the first and second examples previously introduced.

If the property is determined to be a `Measurable Quantity`, it is classified as a `Measure`. Finally, regular expressions are employed to discern whether the property value constitutes a a single value or a range of permissible values, for the purpose of downstream use.

**`Classification` - based on property name**

A set of associated keywords is assembled, and a cosine similarity measure is arrived by pairwise comparison of the property name with each keyword, considering both local and contextual similarity. The maximum pairwise similarity is then compared to a threshold (`0.7`). If it exceeds the threshold, the property is of type `Classification`.

If the property is of either type of `Measure` or `Classification`, it is classified under `Behaviour`.

**`Structure` - based on property name**

For a property to be classified under `Structure`, a check is performed to see if it is associated with the physical, material composition of its associated `Component`. The same custom NER model used in the classification of `Measure` has been trained to classify whether the property name is `Material` or not. If it is, the property is classified under `Structure`.

**Function - based on property name**

A property is classified as of type `Function` if it is either

a. `Application` or

b. `Regulation`

A set of keywords is associated with each, and a cosine similarity measure arrived by pairwise comparison of the property name with each keyword, considering both local and contextual similarity. The maximum pairwise similarity is then compared to a threshold (`0.7`). If it exceeds the threshold for either `Application` or `Regulation,` the property is classified as a `Function`.

Each such successfully classified property is added to a list corresponding to its assigned class. If a property fails to satisfy the conditions for F, B and S, it is classified under `Unknown`.

At the end of this processing step, a dictionary of property lists, one for each facet of FBS, is returned as output.